\documentclass[sort&compress,
    ,final            
  ]
  {aipproc}

\layoutstyle{6x9}

\newcommand{\half}{{\textstyle{\frac{1}{2}}}}
\newcommand{\quarter}{{\textstyle{\frac{1}{4}}}}
\newcommand{\case}[2]{\mbox{$\frac{#1}{#2}$}}

\begin{document}

\title{Hadron Physics and the Dyson--Schwinger Equations of QCD}


\author{Pieter Maris}{
  address={Dept. of Physics and Astronomy, 
 University of Pittsburgh, Pittsburgh, PA 15260},
}

\begin{abstract}
We use the Bethe--Salpeter equation in rainbow-ladder truncation to
calculate the ground state mesons from the chiral limit to
bottomonium, with an effective interaction that was previously fitted
to the chiral condensate and pion decay constant.  Our results are in
reasonable agreement with the data, as are the vector and pseudoscalar
decay constants.  The meson mass differences tend to become constant
in the heavy-quark limit.  We also present calculations for the pion
and rho electromagnetic form factors, and for the single-quark form
factors of the $\eta_c$ and $J/\psi$.
\end{abstract}

\maketitle

\section{Introduction}

Hadrons are color-singlet bound states of quarks, antiquarks, and
gluons.  Bound states appear as poles in the $n$-point functions of a
quantum field theory.  Thus a study of the poles in the $n$-point
functions of QCD will tell us something about hadrons.

In the ultraviolet region, these $n$-point functions can be calculated
using perturbation theory.  For hadronic observables however, we need
to understand the nonperturbative, infrared behavior of the $n$-point
functions of QCD.  The Dyson--Schwinger equations [DSEs], which are
the equations of motion of a quantum field theory, provide us with a
tool to study the $n$-point functions nonperturbatively.  For reviews
on the DSEs and their use in hadron physics,
see~\cite{Roberts:1994dr,Roberts:2000aa,Alkofer:2000wg,Maris:2003vk,Fischer:2006ub}.

\section{Meson physics}

Mesons can be described by solutions of the homogeneous
Bethe--Salpeter equation
\begin{eqnarray}
 \Gamma(p_{\rm out}, p_{\rm in};P) &=& 
        \int\frac{d^4k}{(2\pi)^4} \, K(p_{\rm out},p_{\rm in};k_{\rm out},k_{\rm in}) \,
	\chi(k_{\rm out}, k_{\rm in};P) \, ,
\label{Eq:genBSE}
\end{eqnarray}
with $p_{\rm in}$, $p_{\rm out}$ the 4-momenta of the quark and
antiquark, subject to momentum conservation: $p_{\rm in}-p_{\rm out}=
P$, $\Gamma$ the Bethe--Salpeter amplitude [BSA], and $\chi(k_{\rm
out},k_{\rm in}; P) = S(k_{\rm out}) \, \Gamma(k_{\rm out},k_{\rm in};
P) \, S(k_{\rm in})$; the kernel $K$ is the $q \bar{q}$ scattering
kernel.  This integral equation has solutions $\Gamma$ at discrete
values of $P^2=-M^2$ (in Euclidean metric) of the total meson
4-momentum $P$.  Different types of mesons, such as pseudoscalar or
vector mesons, are characterized by different Dirac structures.  The
properly normalized BSA $\Gamma(p_{\rm out},p_{\rm in};P)$ completely
describes the meson as a $q\bar{q}$ bound state.

Since Eq.~(\ref{Eq:genBSE}) has solutions at discrete values of
$P^2=-M_i^2$, one does not obtain the ``complete'' spectrum, including
the excited states, by solving a matrix equation once; instead, one
has to repeatedly solve Eq.~(\ref{Eq:genBSE}) at different values of
$P^2$ in order to find the mass spectrum.  The ground state in any
particular spin-flavor channel corresponds to the solution with the
lowest mass, $M_0$.  Excited states can be found by looking for
solutions of Eq.~(\ref{Eq:genBSE}) with a larger mass $M_i> M_0$, and
this can indeed be done~\cite{Holl:2004fr,Holl:2005vu}.

\subsection{Rainbow-ladder truncation}

A viable truncation of the infinite set of DSEs has to respect
relevant (global) symmetries of QCD such as chiral symmetry, Lorentz
invariance, and renormalization group invariance.  Here we use the
so-called rainbow-ladder truncation, in which the $q\bar{q}$
scattering kernel is replaced by an effective one-gluon exchange
\begin{eqnarray}
 K(p_{\rm out},p_{\rm in};k_{\rm out},k_{\rm in}) &\to& 
        -4\pi\,\alpha(q^2)\, D_{\mu\nu}^{\rm free}(q)
        \textstyle{\frac{\lambda^i}{2}}\gamma_\mu \otimes
        \textstyle{\frac{\lambda^i}{2}}\gamma_\nu \,,
\end{eqnarray}
where $q=p_{\rm out}-k_{\rm out}=p_{\rm in}-k_{\rm in}$, and
$\alpha(q^2)$ is an effective running coupling.  The corresponding
truncation of the quark DSE is
\begin{eqnarray}
S(p)^{-1}  &=&  i \not\!p\, Z_2 + m_q(\mu)\,Z_4 
 + \case{4}{3}\int\frac{d^4k}{(2\pi)^4} \,4\pi\alpha(q^2)\,
        D_{\mu \nu}^{\rm free}(q) \; \gamma_\mu \, S(k) \, \gamma_\nu  \,,
\label{Eq:quarkDSE}
\end{eqnarray}
where $S(p) = Z(p^2)/[ i \not\!\! p + M(p^2)]$ and $q=k-p$.  This
truncation is the first term in a systematic
expansion~\cite{systematicexp} of the quark-antiquark scattering
kernel $K$; asymptotically, it reduces to leading-order perturbation
theory.  Furthermore, these two truncations are mutually consistent in
the sense that the combination produces vector and axial-vector
vertices satisfying their respective Ward identities.

For the effective interaction we use the 2-parameter model of
Ref.~\cite{Maris:1999nt}
\begin{eqnarray}
\label{gvk2}
\frac{{4\pi\alpha}(q^2)}{k^2} &=&
        \frac{4\pi^2\, D \,k^2}{\omega^6} \, {\rm e}^{-k^2/\omega^2}
        + \frac{ 4\pi^2\, \gamma_m \; {\cal F}(k^2)}
  {\half\ln\Big[{\rm e}^2-1 + 
             \big(1 + k^2/\Lambda_{\rm QCD}^2\big)^2\Big]} \;,
\label{Eq:modelalpha}
\end{eqnarray}
with \mbox{${\cal F}(s)=(1 - {\rm e}^{-s})/s$},
\mbox{$\gamma_m=12/(33-2N_f)$}, and fixed parameters \mbox{$N_f=4$}
and \mbox{$\Lambda_{\rm QCD} = 0.234\,{\rm GeV}$}.  The remaining
parameters, $\omega = 0.4~{\rm GeV}$ and $D = 0.93~{\rm GeV}^2$, were
fitted in~\cite{Maris:1999nt} to reproduce a chiral condensate of
$(240~{\rm MeV})^3$ and $f_\pi=131~{\rm MeV}$.  The first term in
Eq.~(\ref{Eq:modelalpha}) models the infrared enhancement of the
effective $q\bar{q}$ scattering kernel necessary to generate the
experimentally observed amount of dynamical chiral symmetry
breaking~\cite{Hawes:1998cw} .  It was introduced
in~\cite{Maris:1999nt} as a finite-width representation of a
$\delta$-function~\cite{Maris:1997tm}, which can be interpreted as a
regularized $1/p^4$ singularity in
$K$~\cite{McKay:1996th,Alkofer:2006gz}.  The second term ensures the
correct perturbative behavior in the ultraviolet region.

\subsection{Meson spectroscopy}

In Table~\ref{Tab:masses} we give our results for the equal-mass
ground states in each spin channel.  The masses of the light quarks
where fitted in~\cite{Maris:1999nt} to the pion mass (using equal $u$
and $d$ quark masses) and to the kaon mass.  The light vector and
pseudoscalar mesons are described very well by this model: not only
their masses, but also a wide range of other observables agree with
experiments, without adjusting any of the parameters,
see~\cite{Maris:2003vk} and references therein.  Here we apply this
model to heavy quarks as well, and use the vector mesons $J/\psi$ and
$\Upsilon$ to fix the $c$ and $b$ masses.

\begin{table}[b]
\begin{tabular}{l|ll|ll|ccc}
\hline
quark flavor & $M_{\rm PS}$ & $f_{\rm PS}$ 
             & $M_{\rm V}$  & $f_{\rm V}$  
             & $M(0^{++})$ & $M(1^{+-})$ & $M(1^{++})$ \\
\hline
up/down & $0.1385$ & $0.131$ & $0.743$ & $0.207        $ 
             & $0.672$ & $0.83$ & $0.91$ \\
expt.   & $0.135$,$0.140$ & $0.131$ & $0.775$ & $0.221 $ 
             & $0.985$ & $1.23$ & $1.23$ \\ \hline
strange & $0.697 $ & $0.183$ & $1.076$ & $0.260        $ 
             & $1.081$ & $1.17$ & $1.25$ \\
expt.   &  ---     &  ---    & $1.020$ & $0.229     $ 
             &  ---  & --- & ---  \\ \hline
charm   & $2.908 $ & $0.381$ & $3.098$ & $0.421        $ 
             & $3.250$ & $3.26$ & $3.33$ \\
expt.   & $2.980 $ & $0.335\pm0.075$ & $3.097$ & $0.416$ 
             & $3.415$ &   & $3.51$\\ \hline
bottom  & $9.38  $ & $0.66 $ & $9.46 $ & $0.62         $ 
             & $9.72$  & $9.73$ & $9.75$\\
expt.   & $9.30  $ & $     $ & $9.46 $ & $0.715 $ 
             & $9.86$  &   & $9.89$\\
\hline
\end{tabular}
\caption{Masses and leptonic decay constants for equal-mass ground
state $J^{PC}$ mesons.  Experimental data are from
Ref.~\cite{Yao:2006px}, with the exception of
$f_{\eta_c}$~\cite{Edwards:2000bb}. \label{Tab:masses}}
\end{table}
The mass splitting between the pseudoscalar and vector mesons is too
large for the heavy quarkonium states, but the decay constants are in
reasonable agreement with available data.  On the other hand, the mass
splitting between the vector and the scalar mesons is too small; and
the scalar-pseudoscalar mass difference is reasonable.  Also the
axialvector masses are too small, but the mass difference between the
scalar and the $1^{++}$ states is in agreement with data, both for the
light and for the $c$ and $b$ quarks.  Similar results for the light
quark sector and for the charmonium states were found in
Ref.~\cite{Alkofer:2002bp} with a slightly different model
interaction.  Presumably corrections beyond ladder truncation are
necessary for the scalar and axialvector masses: there are significant
cancellations between these corrections in the pseudoscalar and vector
channels~\cite{systematicexp}, but not necessarily in the scalar and
axialvector channels.

\begin{figure}
  \includegraphics[width=.45\textwidth]{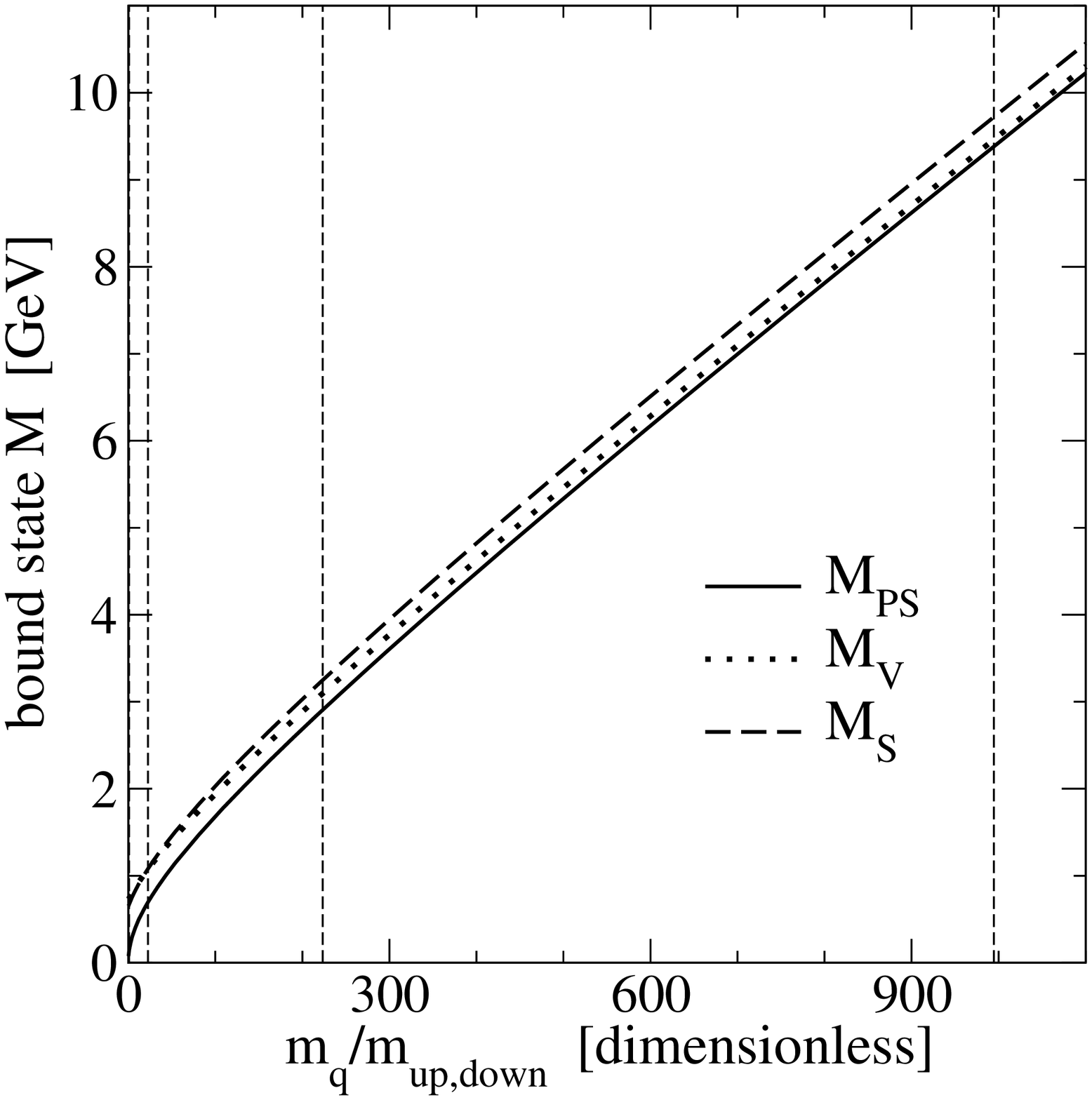}
  \includegraphics[width=.45\textwidth]{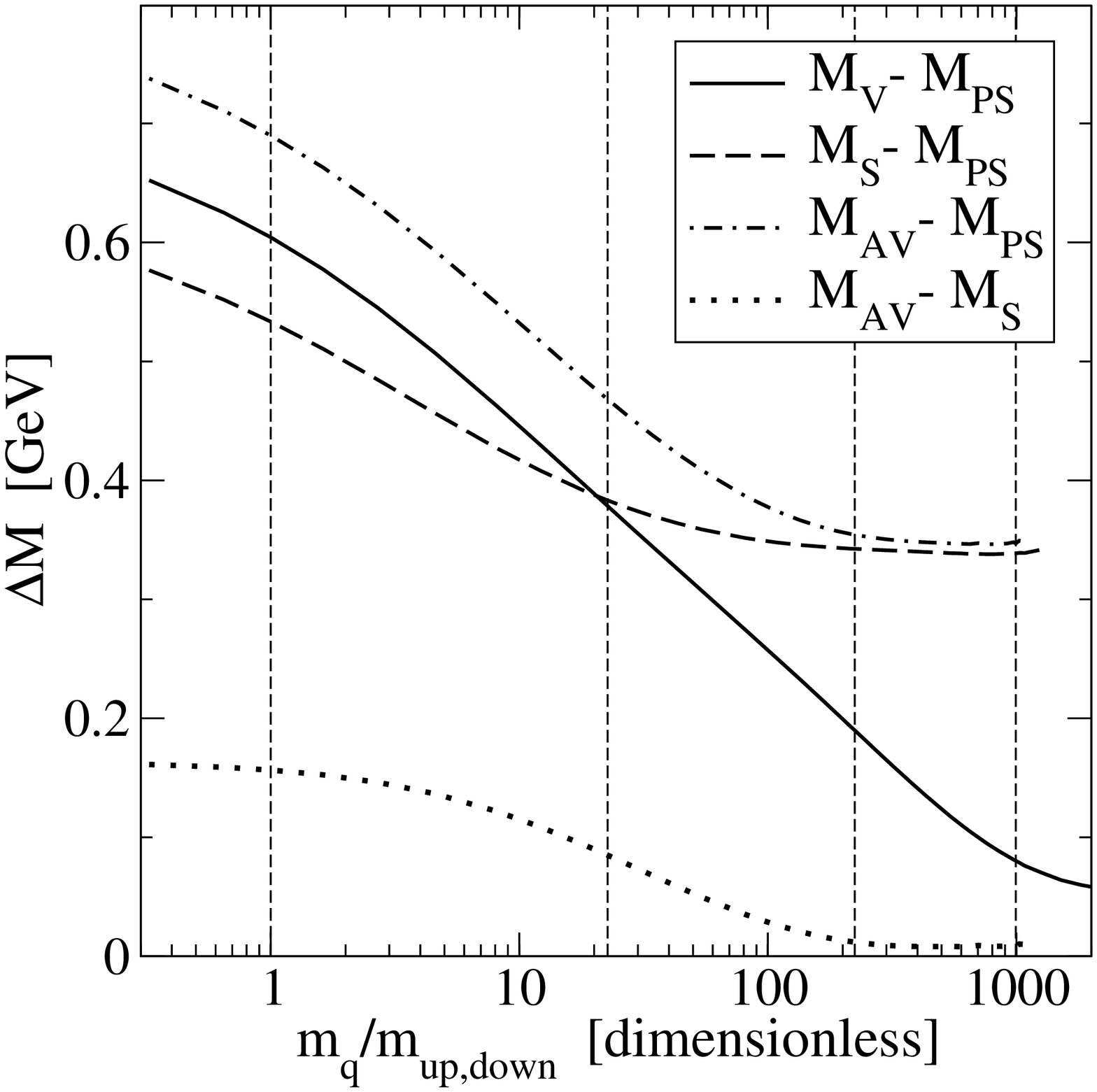}
  \caption{Meson masses (left) and mass differences (right) as
    function of current quark mass, normalized to the up and down
    quark masses.  The vertical dashed lines indicate physical quark
    masses. \label{Fig:masses}}
\end{figure}
Over the entire mass range from the chiral limit up to the bottomonium
states, the pseudoscalar, vector, and scalar masses can be fitted by 
\begin{eqnarray}
  M_{\rm meson}^2 &=& C_0 + C_1 \; m_q + C_2 \; m_q^2 \;,
\end{eqnarray}
where $m_q$ is the current quark mass at our renormalization point
$\mu = 19~{\rm GeV}$.  The fit parameters are $C_0=0$ and $C_1=6.94$
for the pseudoscalars, $C_0=0.51$ and $C_1=7.27$ for the vectors, and
$C_0=0.38$ and $C_1=8.65$ for the scalar mesons, with a common
parameter $C_2 \approx 4.6$.  The fact that the trajectories can all
be fitted with (approximately) the same value for $C_2$ means that for
large masses, the meson mass differences become constant: in the limit
$m_q \to \infty$ the above fit suggests $\Delta M \to \half \Delta
C_1/\sqrt{C_2}$.  Thus this global fit indicates that the mass
difference $M_{\rm V}-M_{\rm PS}$ approaches $0.07~{\rm GeV}$, whereas
$M_{\rm S}-M_{\rm PS}$ approaches $0.4~{\rm GeV}$ for heavy quarks;
our numerical results however do not exclude that the coefficients
$C_1$ are identical for the pseudoscalar and vector mesons, in which
case this mass difference vanishes in the heavy quark limit.

This is indeed consistent if we look at the actual mass differences we
find, see the right panel of Fig.~\ref{Fig:masses}: the mass
difference $M_{\rm V}-M_{\rm PS}$ decreases with increasing quark
mass, it is about $\Delta_M \approx 0.06~{\rm GeV}$ for at $2\,m_b$,
and still decreasing.  Similarly, the mass difference $M_{\rm
AV}-M_{\rm S}$ appears to vanish in the heavy quark limit, but the
differences $M_{\rm S}-M_{\rm PS}$ and $M_{\rm AV}-M_{\rm PS}$ clearly
remain nonzero and appear to go to a constant of about $\Delta M
\approx 0.35~{\rm GeV}$.  However, one should keep in mind that the
model was fitted to the pion decay constant and the chiral condensate;
implicitly we may have incorporated corrections beyond the ladder
kernel in our model for the effective $q\bar{q}$ scattering kernel.
Higher-order corrections affect light quarks differently than heavy
quarks~\cite{Bhagwat:2004hn}.

\begin{table}[b]
\begin{tabular}{lll|cccc}
\hline
$m_q(19)$ & $m_q(2)$ & $m_q(m_q)$ & $M_q\big(p^2=M_q(p^2)^2\big)$ 
          & $M_q(p^2=4)$ & $M_q(p^2=0)$ & $M_q\big(p^2=-\quarter M_V^2\big)$ \\
\hline
\multicolumn{3}{l|}{chiral limit} 
                      & 0.392 & 0.010 & 0.477 & 0.594 \\
0.0037 & 0.005 &      & 0.401 & 0.017 & 0.499 & 0.610 \\
0.0838 & 0.118 &      & 0.556 & 0.168 & 0.689 & 0.845 \\
0.827  & 1.17  & 1.30 & 1.42  & 1.31  & 1.61  & 2.00  \\
3.68   & 5.65  & 4.46 & 4.30  & 4.46  & 4.52  & 5.33  \\
\hline
\end{tabular}
\caption{Current quark masses $m_q(\mu)$ at $\mu=19~{\rm GeV}$, 
scaled down to $\mu=2~{\rm GeV}$ and to $\mu=m_q$ using one-loop pQCD,
together with the dynamical mass function $M(p^2)$ at several values of $p^2$.
\label{Tab:qrkmss}}
\end{table}
\begin{figure}
  \includegraphics[width=.63\textwidth]{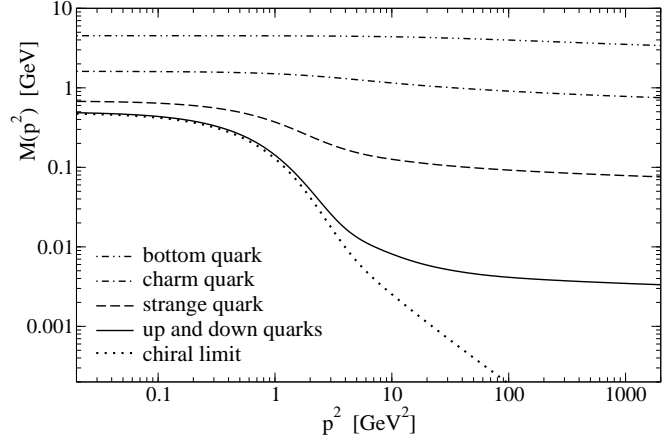}
  \caption{Dynamical quark mass function $M_q(p^2)$ for $u=d$, $s$,
    $c$, $b$ and chiral quarks. 
  \label{Fig:qrkmass}}
\end{figure}
The corresponding quark mass functions are shown in
Fig.~\ref{Fig:qrkmass}, and summarized in Table~\ref{Tab:qrkmss}.  Our
current quark masses are in good agreement with conventional
values~\cite{Yao:2006px} of both the light and the heavy quark masses.
For the light quarks, the nonperturbative mass function $M_q(p^2)$ is
significantly larger than the perturbative quark mass $m_q(\mu)$ at
$p=2=\mu$, indicating that chiral symmetry breaking sets in well above
this scale.  The momentum dependence of $M_{c,b}(p^2)$ is much less
dramatic.  Nevertheless, there is a significant difference between the
dynamical mass in the region relevant for $q\bar{q}$ bound states,
namely $p^2 \sim -\quarter M_{\rm meson}^2$ in the timelike region,
and $M_{c,b}(p^2)$ in the spacelike region, even for $b$ quarks.  For
$0 < p^2 < -\quarter M_{\rm meson}^2$, the mass function of the heavy
quarks is in fact quite close to the typical pole masses used in
non-relativistic calculations of charmonium, $m_c^{\rm pole} \approx
1.47$ to $1.83$ GeV and bottomonium, $m_b^{\rm pole} \approx 4.7$ to
$5.0$ GeV~\cite{Yao:2006px}.

\subsection{Electromagnetic form factors}

The $q\bar{q}\gamma$ vertex is the solution of the renormalized
inhomogeneous Bethe--Salpeter equation with the same kernel $K$ as
Eq.~(\ref{Eq:genBSE}).  Thus for photon momentum $Q$, we have
\begin{eqnarray}
\Gamma_\mu(p_{\rm out},p_{\rm in}) = Z_2\, \gamma_\mu + \int\frac{d^4k}{(2\pi)^4} 
        K(p_{\rm out},p_{\rm in};k_{\rm out},k_{\rm in}) \; S(k_{\rm out})\,
        \Gamma_\mu(k_{\rm out},k_{\rm in}) \, S(k_{\rm in}) \,,
\label{Eq:vectorBSE}
\end{eqnarray}
with $p_{\rm out}$ and $p_{\rm in}$ the outgoing and incoming quark momenta,
respectively, and similarly for $k_{\rm out}$ and $k_{\rm in}$, with
$p_{\rm out}-p_{\rm in}=k_{\rm out}-k_{\rm in}=Q$.  The ladder truncation for
Eq.~(\ref{Eq:vectorBSE}), in combination with the rainbow truncation
for the quark propagators and impulse approximation for
electromagnetic form factors, satisfies the vector Ward--Takahashi
identity and electromagnetic current conservation is guaranteed.

Also note that solutions of the {\em homogeneous} version of
Eq.~(\ref{Eq:vectorBSE}) define vector meson bound states with masses
\mbox{$M_{\rm V}^2=-Q^2$} at discrete timelike momenta $Q^2$.  It
follows that $\Gamma_\mu$ has poles at those locations.  Thus the
effects of intermediate vector meson states on electromagnetic
processes can be unambiguously incorporated by using the properly
dressed $q\bar{q}\gamma$ vertex rather than the bare vertex
$\gamma_\mu$~\cite{Maris:1999bh}.

Consider for example the 3-point function describing the coupling of a
photon with momentum $Q$ to the quark $a$ of a meson $a\bar{b}$, with
initial and final momenta \mbox{$P \pm \half Q$}
\begin{eqnarray}
\Lambda^a_\mu(P,Q) &=&
        i\,N_c \int\frac{d^4k}{(2\pi)^4} {\rm Tr}\big[ \Gamma^{a}_\mu(q_-,q_+)
	  \, \chi^{a\bar{b}}(q_+,q) \;
        S^b(q)^{-1} \, \bar\chi^{\bar{b}a}(q,q_-)\big] \;,
\label{Eq:triangle}
\end{eqnarray}
with \mbox{$q = k-\half P$} and \mbox{$q_\pm = k+\half P \pm \half Q$}.  The
corresponding single-quark elastic form factor $F^a$ of a pseudoscalar
meson is defined by
\begin{eqnarray}
  2\;P_\mu\;F^a(Q^2) &=& \Lambda^a_\mu(P,Q) \,.
\end{eqnarray}
Vector mesons have three elastic form factors, commonly referred to as
the electric, magnetic, and quadrupole form factors $G_{\rm E}(Q^2)$,
$G_{\rm M}(Q^2)$, and $G_{\rm Q}(Q^2)$.  The electric monopole moment
(i.e. the electric charge), magnetic dipole moment and the electric
quadrupole moment follow from the values of these form factors in the
limit $Q^{2}\to 0$: $G_{\rm E}(0) = 1$ (constrained by current
conservation), $G_{\rm M}(0) = \mu$, and $G_{\rm Q}(0)= {\cal Q}$.

\begin{table}[t]
\begin{tabular}{l|c|cccc}
   & \multicolumn{1}{c|}{$r^2_{\rm PS}$} 
   & \multicolumn{1}{c}{$r^2_{\rm V,E}$}
   & \multicolumn{1}{c}{$\mu$}
   & \multicolumn{1}{c}{$r^2_{\rm V,M}$}
   & \multicolumn{1}{c}{$\cal Q$}  \\ 
\hline
up/down & $0.44$  & $0.54$  & $2.01$ & $0.49$ & $-0.41$ \\
\hline
charm   & $0.048$ & $0.052$ & $2.13$ & $0.047$ & $-0.28$ \\
lattice~\cite{Dudek:2006ej}
        & 0.063(1)& 0.066(2) &  2.10(3) & & -0.23(2) \\
\hline
\end{tabular}
\caption{Static electromagnetic properties of pseudoscalar and vector
  $u\bar{d}$ mesons ($\pi$ and $\rho$) and $c\bar{c}$ mesons (fictitious). 
  \label{Tab:EMF}}
\end{table}
\begin{figure}
  \includegraphics[width=.45\textwidth]{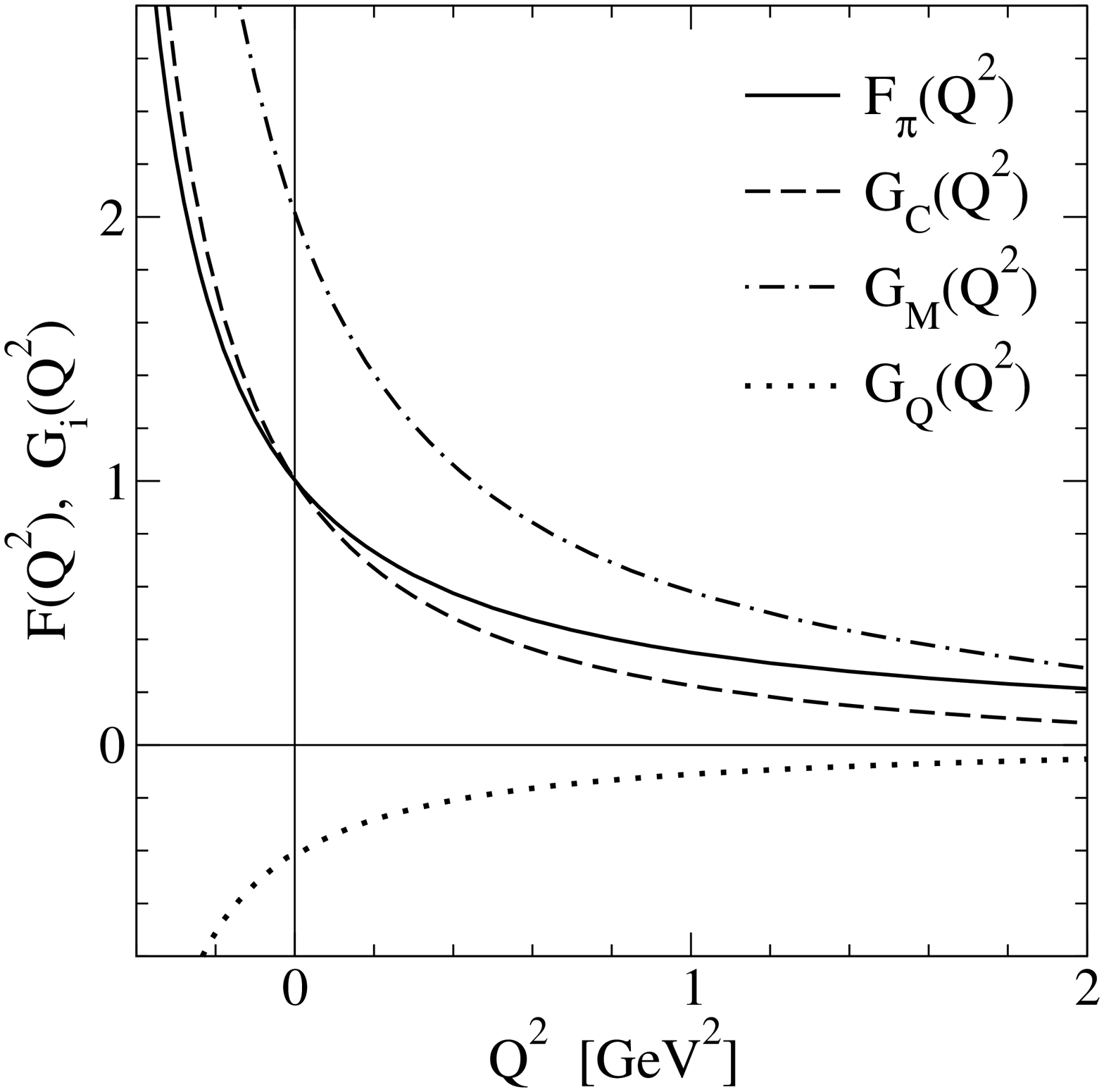}
  \includegraphics[width=.45\textwidth]{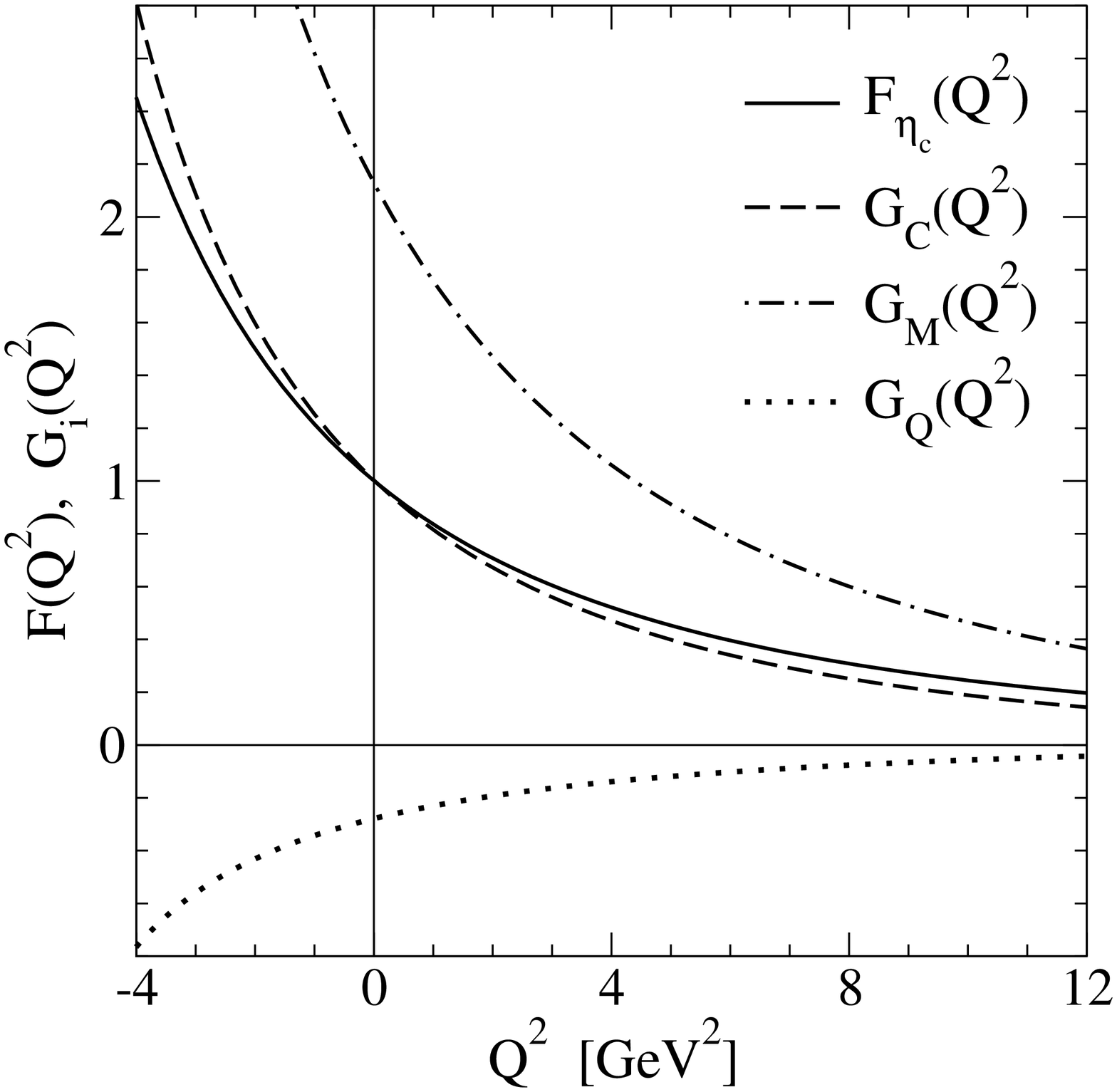}
  \caption{Single-quark form factors: $\pi$ and $\rho$ (left)
    and $c\bar{c}$ pseudoscalar and vector mesons (right).
    \label{Fig:EMF}}
\end{figure}
Our results for the pion form factor~\cite{Maris:1999bh,Maris:2000sk}
are in good agreement with the data, both in the spacelike
region~\cite{Fpinewexp} and in the timelike region; the charge radius
agrees very well with the experimental value $\langle r_\pi^2 \rangle
= 0.44\pm0.01~{\rm fm}^2$~\cite{A86}, see Table~\ref{Tab:EMF}.  The
vector charge radius~\cite{BhagwatNew} is slightly larger than the
pseudoscalar radius, both for light quarks and for charm quarks.  This
suggests that the vector states are broader than the corresponding
pseudoscalar states, assuming that the charge distribution is
indicative of the physical size of the bound state.  This agrees with
the naive intuition that a more tightly bound state is more compact
than a heavier state with the same constituents.  For charm quarks
this difference is significantly smaller than for up and down quarks,
in agreement with recent lattice calculations~\cite{Dudek:2006ej}.
The magnetic moment appears to be surprisingly independent of the
quark mass; the quadrupole moment decreases with increasing quark
mass~\cite{BhagwatNew}.  Recent lattice
simulations~\cite{Dudek:2006ej} agree quite well with our results for
the moments of the single-quark form factors of the $J/\psi$.

In Fig.~\ref{Fig:EMF} we see that both the pseudoscalar $F_\pi$ and
all three vector form factors $G_i^\rho$ diverge in the timelike
region as $Q^2 \to -0.55~{\rm GeV}^2$, corresponding to the
vector-meson poles in the dressed quark photon vertex.  Similarly, the
single-quark form factors of the $\eta_c$ and $J/\psi$ diverge as $Q^2
\to -9.5~{\rm GeV}^2$.  However, it is only the pion form factor that
can be described by a vector meson dominance [VMD] curve,
$F_\pi\approx M_\rho^2/[Q^2 + M_\rho^2]$, over the entire $Q^2$-region
shown.  The $\rho$ form factors $G_i^\rho$ drop significantly
faster~\cite{BhagwatNew} than a VMD curve, as do the $c\bar{c}$ form
factors, both for pseudoscalar and vector states.


\begin{theacknowledgments}
This work was supported by the US Department of Energy, contract
No.~DE-FG02-00ER41135, and benefited from the facilities of the NSF
Terascale Computing System at the Pittsburgh Supercomputing Center.
\end{theacknowledgments}



\begin{thebibliography}{99}

\bibitem{Roberts:1994dr}
C.D.~Roberts and A.G.~Williams,
Prog.\ Part.\ Nucl.\ Phys.\  {\bf 33}, 477 (1994)
[arXiv:hep-ph/9403224].

\bibitem{Roberts:2000aa}
C.D.~Roberts and S.M.~Schmidt,
Prog.\ Part.\ Nucl.\ Phys.\  {\bf 45S1}, 1 (2000)
[arXiv:nucl-th/0005064].

\bibitem{Alkofer:2000wg}
R.~Alkofer and L.~von Smekal,
Phys.\ Rept.\  {\bf 353}, 281 (2001)
[arXiv:hep-ph/0007355].

\bibitem{Maris:2003vk}
P.~Maris and C.D.~Roberts,
Int.\ J.\ Mod.\ Phys.\ E {\bf 12}, 297 (2003)
[arXiv:nucl-th/0301049].

\bibitem{Fischer:2006ub}
C.S.~Fischer,
J.\ Phys.\ G {\bf 32}, R253 (2006)
[arXiv:hep-ph/0605173].

\bibitem{Holl:2004fr}
A.~Holl, A.~Krassnigg and C.D.~Roberts,
Phys.\ Rev.\ C {\bf 70}, 042203 (2004)
[arXiv:nucl-th/0406030].

\bibitem{Holl:2005vu}
A.~Holl {\it et al.},
Phys.\ Rev.\ C {\bf 71}, 065204 (2005)
[arXiv:nucl-th/0503043].

\bibitem{systematicexp}
A.~Bender, C.D.~Roberts and L.~Von Smekal,
Phys.\ Lett.\ B {\bf 380}, 7 (1996)
[arXiv:nucl-th/9602012];
A.~Bender {\it et al.},
Phys.\ Rev.\ C {\bf 65}, 065203 (2002)
[arXiv:nucl-th/0202082].

\bibitem{Bhagwat:2004hn}
M.S.~Bhagwat, A.~Holl, A.~Krassnigg, C.D.~Roberts and P.C.~Tandy,
Phys.\ Rev.\ C {\bf 70}, 035205 (2004)
[arXiv:nucl-th/0403012].

\bibitem{Maris:1999nt}
P.~Maris and P.C.~Tandy,
Phys.\ Rev.\ C {\bf 60}, 055214 (1999)
[arXiv:nucl-th/9905056].

\bibitem{Hawes:1998cw}
F.T.~Hawes, P.~Maris and C.D.~Roberts,
Phys.\ Lett.\ B {\bf 440}, 353 (1998)
[arXiv:nucl-th/9807056].

\bibitem{Maris:1997tm}
P.~Maris and C.D.~Roberts,
Phys.\ Rev.\ C {\bf 56}, 3369 (1997)
[arXiv:nucl-th/9708029].

\bibitem{McKay:1996th}
D.W.~McKay and H.J.~Munczek,
Phys.\ Rev.\ D {\bf 55}, 2455 (1997)
[arXiv:hep-th/9607075].

\bibitem{Alkofer:2006gz}
R.~Alkofer, C.S.~Fischer and F.J.~Llanes-Estrada,
arXiv:hep-ph/0607293.

\bibitem{Yao:2006px}
W.~M.~Yao {\it et al.}  [Particle Data Group],
J.\ Phys.\ G {\bf 33}, 1 (2006).

\bibitem{Edwards:2000bb}
K.W.~Edwards {\it et al.}  [CLEO Collaboration],
Phys.\ Rev.\ Lett.\  {\bf 86}, 30 (2001)
[arXiv:hep-ex/0007012].

\bibitem{Alkofer:2002bp}
R.~Alkofer, P.~Watson and H.~Weigel,
Phys.\ Rev.\ D {\bf 65}, 094026 (2002)
[arXiv:hep-ph/0202053].

\bibitem{Maris:1999bh}
P.~Maris and P.C.~Tandy,
Phys.\ Rev.\ C {\bf 61}, 045202 (2000)
[arXiv:nucl-th/9910033].

\bibitem{Maris:2000sk}
P.~Maris and P.C.~Tandy,
Phys.\ Rev.\ C {\bf 62}, 055204 (2000)
[arXiv:nucl-th/0005015].

\bibitem{Fpinewexp}
V.~Tadevosyan {\it et al.}  [Fpi-1 Collaboration],
arXiv:nucl-ex/0607007;
T.~Horn {\it et al.}  [Fpi2 Collaboration],
arXiv:nucl-ex/0607005.

\bibitem{A86} 
S.~R.~Amendolia {\it et al.}  [NA7 Collaboration],
Nucl.\ Phys.\ B {\bf 277}, 168 (1986).

\bibitem{BhagwatNew}
M.S.~Bhagwat and P.~Maris,
in preparation.

\bibitem{Dudek:2006ej}
J.J.~Dudek, R.G.~Edwards and D.G.~Richards,
Phys.\ Rev.\ D {\bf 73}, 074507 (2006)
[arXiv:hep-ph/0601137].

\end{thebibliography}
\end{document}